\documentstyle[prl,aps,twocolumn,psfig]{revtex}

\draft
\begin{document}
\twocolumn[\hsize\textwidth\columnwidth\hsize\csname @twocolumnfalse\endcsname
\title{Griffiths phase manifestation in disordered dielectrics}
\author{{\bf {V.A.Stephanovich}}}
\address{Institute of Mathematics, University of Opole, Oleska 48, 45-052, Opole, Poland\\
and Institute of Semiconductor Physics NAS of Ukraine, Kiev, Ukraine}
\date{\today}
\maketitle
\begin{abstract}
We predict the existence of Griffith phase in the dielectrics with
concentrational crossover between dipole glass
(electric analog of spin glass) and
ferroelectricity. The peculiar representatives of above substances are
$KTaO_3:Li$, $Nb$, $Na$ or relaxor ferroelectrics like
$Pb_{1-x}La_xZr_{0.65}Ti_{0.35}O_3$.

Since this phase exists above ferroelectric phase
transition temperature (but below that temperature for ordered substance),
we call it "para-glass phase". We assert that the
difference between paraelectric and para-glass phase of above substances is
the existence of clusters (inherent to "ordinary" Griffiths phase in Ising
magnets) of correlated dipoles. We show that randomness play a decisive
role in
Griffiths (para-glass) phase formation; this phase does not exist in a mean
field approximation.

To investigate Griffiths phase properties, we calculate
the density of Yang-Lee (YL) zeros of  partition function and found
that it has "tails" inherent to Griffith phase in the above temperature
interval.

We perform calculations on the base of our self-consistent
equation for long-range order parameter in the external electric field.
This equation has been derived in the random field theory framework.
Latter theory automatically incorporates both short-range (due to
indirect interaction via transversal optical phonons of the
host lattice) and long-range (ordinary dipole-dipole interaction)
interactions between impurity dipoles so that the problem of
long-range interaction consideration does not appear in it.

\end{abstract}

]

It was shown by Griffiths \cite{isx} that the free energy of a dilute Ising ferromagnet is
a nonanalytic function of the external magnetic field for all temperatures between
critical value $T_c(x)$ ($x$ is concentration of lattice sites without Ising spins)
and $T_c(1)$ of the corresponding nondilute system. The manifestation of nonanalyticity in
the original article of Griffiths was the distribution of the zeros of the partition function $Z$
(the Yang-Lee (YL) zeros) in the plane of complex magnetic fields $H=i\theta$.
Namely, below $T_c(1)$ YL zeros appear arbitrarily close to the point $H=0$, implying a zero radius of convergence for
expansions of thermodynamic quantities in powers of $H$. The "visible sign" of such
nonanalyticity is the appearance of "tail" in density of YL zeros $\rho(\theta)$ at
$T_c(x)<T<T_c(1)$.

Dielectric systems are, in many respects, different from magnetic ones. The main difference
is that main interaction in dielectrics is long-range dipole-dipole interaction, which is small
relativistic correction to the short-range exchange interaction in magnets. It was shown in
\cite{tim} that former interaction  in disordered dielectrics does not prevent the occurrence
of aforementioned "tails" in  $\rho(\theta)$ and thus to realization of Griffiths phase analog.

In the present paper we investigate the manifestation of Griffiths phase in the dielectrics,
where both long-range ferroelectric order and dipole glass (say, "dielectric spin glass")
phase can occur depending on impurity dipoles concentration. The peculiar representatives
of such substances are $KTaO_3:Li$, $Nb$, $Na$, where $Li$, $Nb$ or $Na$, being
off-center ions, constitute the impurity dipoles with some discrete number of permissible
orientations in host $KTaO_3$ matrix (see \cite{rmf} for details). Other peculiar representatives
of aforementioned class of substances are relaxor ferroelectrics like
$Pb_{1-x}La_xZr_{0.65}Ti_{0.35}O_3$ (PLZT), which can be considered as some
reference phase (in the case of PLZT it is $PbZr_{0.65}Ti_{0.35}O_3$
having long-range ferroelectric order), "spoiled" by intrinsic defects of differnt kind,
playing a role of impurity dipoles (see \cite{glfar} for details).

Although the description of aforementioned disordered dielectrics is still not complete
at present, we suggested so-called random local field method (see, e.g. \cite{stef97} and
references therein) for their description.
This method captures remarkably well the peculiarities of physical properties of
aforementioned substances, namely concentrational crossover between
ferroelectric and dipole glass phases, existence of mixed ferro-glass phase and
nonexponential long-time relaxation in glass and ferro-glass phases \cite{stef97}.

In this method, we derive the
equation for long-range order parameter $L$. This equation was derived self-consistently
through the distribution function of random fields, acting between impurity dipoles (see
\cite{stef97} for details). The specific form of this equation depends on particular number
of  orientations of impurity dipole in the host crystal matrix. For the simplest case of the
impurity dipole with only two permissible orientations this equation has the form
\begin{eqnarray}
&&L =kT\int_{-\infty }^{\infty }\tanh \left( \beta E\right) f(E,L)dE,
\nonumber \\
&&f(E,L)=\frac{1}{2\pi }\int_{-\infty }^{\infty }\exp (iE\rho )\times
\nonumber \\
&&\times \exp \left[ F_{1}({\rho })+iF_{2}({\rho })\right] d\rho ,  \nonumber \\
&&F_{1}(\rho ) =n\int_{V}\left[ \cos ({\cal K}_{zz}\rho )-1\right] d^{3}r,
\nonumber \\
&&F_{2}(\rho ) =nL\int_{V}\sin ({\cal K}_{zz}\rho )d^{3}r\equiv nE_{0}(\rho
)L,  \label{dy12}
\end{eqnarray}
where $n$ is impurities concentration,
$f(E,L)$ is a distribution function of random fields (see, e.g. \cite{stef97}),
$F_1(\rho)$ is "responsible" for dispersion of random fields,
$F_2(\rho)$ - for mean value of random field,
${\cal K}_{zz}$ is a component of tensor of interaction
${\cal K}_{\alpha \beta}({\vec r})$ ($\alpha, \beta =x,y,z $,
${\vec r}$ is interimpurity separation)
 between impurity dipoles in highly polarizable dielectric host .
General form of  ${\cal K}_{\alpha \beta}$ reads
(see, e.g. \cite{stef93} and references therein)

\begin{eqnarray}
{\cal K}_{\alpha \beta }({\vec{r}})&=&\frac{d^{\ast 2}}{\varepsilon _{0}r^{3}}%
\Biggl\{ f_{1}(r/r_{c})\delta _{\alpha \beta }+ \nonumber \\
&&+(3m_{\alpha }m_{\beta }-\delta
_{\alpha \beta })(1+f_{2}(r/r_{c}))\Biggr\} ,\nonumber \\
{\vec{m}}&=&\frac{\vec{r}}{r}
\label{dyy1}
\end{eqnarray}

 where $\varepsilon _0$ is host lattice static dielectric permittivity, $d^*$
is the effective dipole moment of impurity (see, e.g. \cite{rmf}),
 $r_{c}$ is host lattice correlation radius (two impurities at a
distance less or equal $r_{c}$ ''feel'' each other). The physical reason for
appearance of $r_{c}$ and functions
$f_{1}(x),  f_{2}(x) \propto x^4\exp(-x)$ is
indirect interaction of impurity dipoles via host lattice soft phonon mode
(see, e.g.,\cite{stef93} and references therein). If latter
interaction is absent, we have $r_{c}\rightarrow 0,\ $ $f_{1}(r/r_{c})%
, f_{2}(r/r_{c})\rightarrow 0$ so that
(\ref{dyy1}) gives ordinary dipole-dipole interaction. It is seen that interaction
(\ref{dyy1}) incorporates both long-range part (term $(3m_{\alpha }m_{\beta }-\delta
_{\alpha \beta })/r^3$) and short-range part (terms containing $f_{1}(x)$ and  $f_{2}(x)$).
The explicit form of the functions $f_{1}$ and $f_{2}$ can be found,
e.g. in \cite{glfar,stef93}.

It can be shown (see, e.g. \cite{stef97}), that at high impurity concentrations distribution function
of random fields has Gaussian form. This form can be obtained from (\ref{dy12}) by the
expansion of integrands in $F_{1,2}(\rho)$ up to first leading order. This gives
\begin{eqnarray}
F_1(\rho)=c\rho ^2, F_2(\rho)=nE_0L, \nonumber \\
c=\frac{16 \pi}{15}\frac{\left(nd^{\ast 2}\right)^2}{(\varepsilon _{0})^2nr_c^3},
E_{0}=4\pi \frac{nd^{\ast 2}}{\varepsilon _{0}}.\label{qe0}
\end{eqnarray}
Since the case of Gaussian distribution function implies the strongest 
(both long-range
and short-range) interaction between impurity dipoles 
(and consequently their clusters)
this case can be regarded as a most difficult for Griffiths phase 
to occur. We shall
demonstrate the manifestation of Griffith phase for this case 
with understanding that for lower
impurity dipoles concentration the realization of Griffiths phase is easier.

The equation for order parameter for Gaussian distribution function 
assumes the form
\begin{eqnarray}
L=\frac{2}{\pi }\int_{0}^{\infty }\int_{0}^{\infty }\tanh \left( \frac{E}{kT}%
\right) \exp (-c\rho ^{2})\times \nonumber \\
\times \sin (\rho E)\sin \rho ({\cal E}+E_{0}L)d\rho dE.
\label{qe1}
\end{eqnarray}
Introduce following dimensionless variables
\begin{eqnarray}
\frac{{\cal E}}{E_{0}}=h,\ \rho E_{0}=y,\ \frac{E}{E_{0}}=x,\ \frac{kT}{E_{0}%
}=\tau ,\nonumber \\
\frac{c}{E_{0}^{2}}=\frac{1}{15\pi z},\ z=nr_{c}^{3}.  \label{qe2}
\end{eqnarray}
In these variables the equation for $L$ assumes the form
\begin{eqnarray}
&&L=\frac{2}{\pi }\int_{0}^{\infty }\int_{0}^{\infty }\tanh \left( \frac{x}{%
\tau }\right) \exp \left( -\frac{y^{2}}{15\pi z}\right) \times \nonumber \\
&&\times \sin \left(xy\right) \sin y\left( h+L\right) dxdy.  \label{qe3}
\end{eqnarray}
Integration over $x$ in (\ref{qe3}) gives
\begin{equation}
\tau \int_{0}^{\infty }\exp \left( -\frac{y^{2}}{15\pi z}\right) \frac{\sin
y(h+L)}{\sinh \left( \frac{\pi y\tau }{2}\right) }dy=L.  \label{qe5}
\end{equation}
We shall investigate the Griffiths phase properties on the base of
Eq (\ref{qe5}).

The equation for phase transition temperature $\tau _c$ can be obtained from
(\ref{qe5}) at
$L\to 0$ (of course, at $h=0$)
\begin{equation}
\tau _{c}\int_{0}^{\infty }y\exp \left( -\frac{y^{2}}{15\pi z}\right) \frac{%
dy}{\sinh \left( \frac{\pi y\tau _{c}}{2}\right) }=1.  \label{qe7}
\end{equation}
The equation for critical concentration of impurities can be obtained
from (\ref{qe7}) at $\tau _{c}\to 0$. It reads

\begin{equation}
\frac{2}{\pi }\int_{0}^{\infty }\exp \left( -\frac{y^{2}}{15\pi z_{cr}}\right)
dy=1,\text{or }z_{cr}=\frac{1}{15}.  \label{qe8}
\end{equation}
Equations (\ref{qe7},\ref{qe8}) determine the phase diagram of the system
under consideration. It is shown on the Fig.1.

\begin{figure}[th]
\vspace*{-5mm}
\centerline{\centerline{\psfig{figure=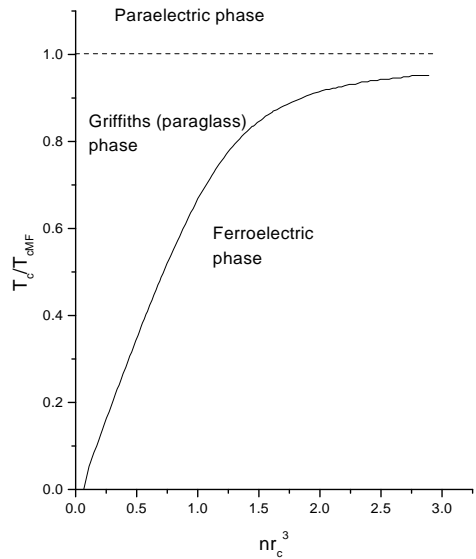,width=0.8\columnwidth}}}
\caption{Phase diagram of the substances under consideration.}
\end{figure}

The theorem of Yang and Lee states that zeros of a partition function $Z$
lie on the unit circle $h=i\theta$ in the plane of complex variable
$z=\exp(-2\beta H)$, ($\beta=1/k_bT$)\cite{ly}. To calculate the density
of YL zeros, we put in (\ref{qe5})
\begin{equation}
h=i\theta ,\ L=L_{1}+iL_{2}.  \label{qe9}
\end{equation}
We have from (\ref{qe5})

\begin{eqnarray}
L_{1} &=&\tau \int_{0}^{\infty }\exp \left( -\frac{y^{2}}{15\pi z}\right)
\frac{\sin (L_{1}y)\cosh \left( y(\theta +L_{2})\right) }{\sinh \left( \frac{%
\pi y\tau }{2}\right) }dy,  \nonumber \\
L_{2} &=&\tau \int_{0}^{\infty }\exp \left( -\frac{y^{2}}{15\pi z}\right)
\frac{\cos (L_{1}y)\sinh (y(\theta +L_{2}))}{\sinh \left( \frac{\pi y\tau }{2%
}\right) }dy.\nonumber \\
\label{qe10}
\end{eqnarray}
The density of YL zeros  $\ \rho (\theta )$ equals \cite{osn}
\begin{equation}
\rho (\theta )=\frac{L_{1}}{\pi }.  \label{qe11}
\end{equation}
The equations (\ref{qe10}),(\ref{qe11}) are the main theoretical result
of this paper. To find $\rho (\theta )$ from (\ref{qe11}), we should
find (at given temperature and
concentration) $L_{1}$ and $L_{2}$ from (\ref{qe10}). Griffits phase
is realized in the substances under consideration if $\rho (\theta )$
has "tails" near $\theta =0$ at $\tau _c<\tau < 1$.
This is because in the "language" of equations (\ref{qe7}),(\ref{qe8}) the
phase transition temperature of nondilute system $T_c(1)$ corresponds
to the quantity $E_0$, determining the phase transition temperature
in a mean field approximation $T_{cMF}$, i.e.
$T_c(1) \equiv E_0=T_{cMF}$.

\begin{figure}[th]
\vspace*{-5mm}
\centerline{\centerline{\psfig{figure=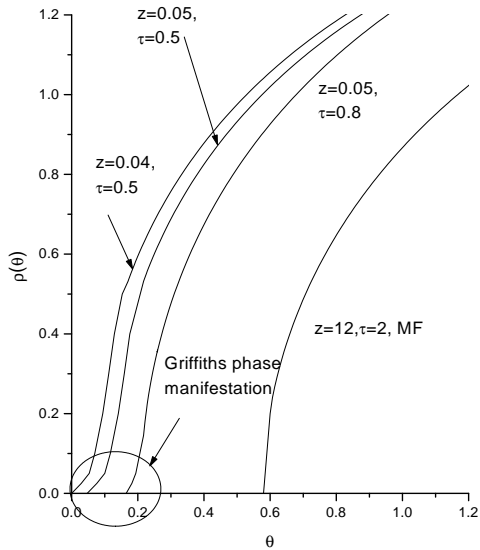,width=0.8\columnwidth}}}
\caption{Density of YL zeros. Characteristic "tails" inherent to Griffith
phase are shown. Curve labeled "MF" is also realized in mean
field approximation (see Eq.(\ref{qe13})).}
\end{figure}

We plot the dependence $\rho (\theta )$ for different $\tau $ and $z$ in the
Fig.2. First of all, it is seen the presence of "tails", inherent to
Griffiths phase for dilute system. For nondilute (completely ordered system)
there is no Griffiths singularities.
This clearly demonstrates the occurrence of Griffiths
phase in the dielectrics with concentrational crossover between dipole glass
and ferroelectricity. Fig.2 permits us to show the region of existence of
Griffiths phase in the phase diagram of substances under consideration.

Since Griffiths phase occurs at temperatures higher than that of
ferroelectric phase transition of disordered system, but lower
than that of the ordered (nondilute) system, we can call
this phase "para-glass".
Glassy behaviour is due to clusters of impurity dipoles
(see \cite{tim,br,sm,osn}). The manifestation
of these clusters is indeed "tails" in $\rho (\theta )$.
It is seen from Fig.2 that at temperatures than $T_{cMF}$
there is no more "tails" in $\rho (\theta )$ and para-glass phase
becomes conventional paraelectric phase.

Other peculiar feature of all curves is the existence of some threshold
value $\theta _e$, below which there are no YL zeros
($\rho (\theta ) \equiv 0$). This is so-called YL edge, which is
manifestation of ordered state of the system (paramagnetic phase of nondilute
system, see \cite{br,sm,osn}). The
"tail-like" approach to $\theta _e$ at smaller $z$ is due to disorder
(glassy effects)
in the system. Thus at sufficiently small $z$
($z=0.05$ in Fig.2) the system is in a mixed ferro-glass phase, exhibiting
both "order" and "disorder" features. The difference between ferro-glass
and para-glass phase is that while in former the overall long-range
order exists in the system, in latter it does not exist. Our supposition
that para-glass phase corresponds to superparamagnetism,
well-known in spin glasses.

To find the equation for $\theta _{e}$ we put $L_{1}=0$ in (\ref{qe10}).
\begin{eqnarray}
L_{2} &=&\tau \int_{0}^{\infty }\exp \left( -\frac{y^{2}}{15\pi z}\right)
\frac{\sinh (y(\theta _{e}+L_{2}))}{\sinh \left( \frac{\pi y\tau }{2}\right)
}dy.  \label{qe21} \\
1 &=&\tau \int_{0}^{\infty }\exp \left( -\frac{y^{2}}{15\pi z}\right) \frac{%
y\cosh (y(\theta _{e}+L_{2}))}{\sinh \left( \frac{\pi y\tau }{2}\right) }dy.
\nonumber
\end{eqnarray}
The dependence $\theta _e (\tau )$ is shown in Fig.3. It is seen that at
$z>z_{cr}=1/15$ (ferroelectric and ferro-glass phase)
$\theta _e=0$ at $\tau =\tau _c$, while at $z<z_{cr}$ $\theta _e \neq 0$
for all temperatures. To trace the transition from ordered to Griffith phase
it is instructive to investigate the concentrational dependence of
$\theta _e$ at zeroth temperature.

\begin{figure}[th]
\vspace*{-5mm}
\centerline{\centerline{\psfig{figure=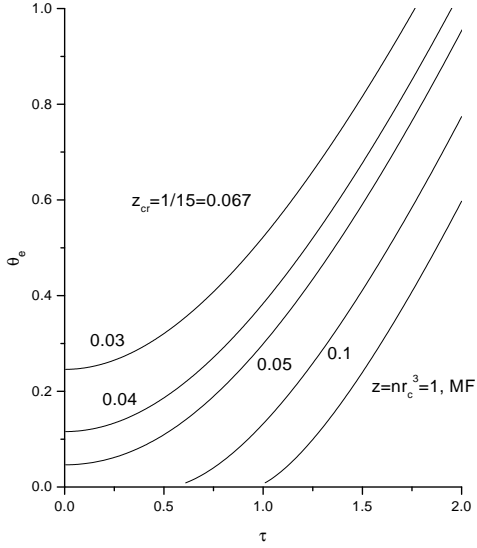,width=0.8\columnwidth}}}
\caption{Temperature dependence of $\theta _e$ at different impurity
concentration. Curve labeled "MF" is also realized in mean
field approximation (see Eq.(\ref{qe17})).}
\end{figure}

We have from (\ref{qe21})
\begin{eqnarray}
L_{2} &=&\frac{2}{\pi }\int_{0}^{\infty }\exp \left( -\frac{y^{2}}{15\pi z}%
\right) \frac{\sinh (y(\theta _{e}+L_{2}))}{y}dy, \label{qq0} \\
1 &=&\frac{2}{\pi }\int_{0}^{\infty }\exp \left( -\frac{y^{2}}{15\pi z}%
\right) \cosh (y(\theta _{e}+L_{2}))dy.\label{qq1}
\end{eqnarray}
Integral in (\ref{qq1}) can be calculated analytically giving
\[
\sqrt{15z}\exp \left[ \frac{15\pi z}{4}(\theta +L_{2})^{2}\right] =1.
\]
Denoting $\mu =\theta +L_{2}$, we obtain
\begin{eqnarray}
\theta _{e} &=&\mu -\frac{2}{\pi }\int_{0}^{\infty }\exp \left( -\frac{y^{2}%
}{15\pi z}\right) \frac{\sinh (\mu y)}{y}dy, \nonumber \\
\mu ^{2} &=&-\frac{2}{15\pi z}\ln (15z).  \label{qq2}
\end{eqnarray}
or
\begin{eqnarray}
&&\theta _{e}(\lambda )=\frac{\lambda }{\sqrt{\pi }}\exp \left( \frac{\lambda
^{2}}{4}\right) -\frac{2}{\pi }\int_{0}^{\infty }\exp (-t^{2})\frac{\sinh
(\lambda t)}{t}dt,\nonumber \\
&&\lambda ^{2}=-2\ln (15z).  \label{qe28}
\end{eqnarray}
The dependence (\ref{qe28}) is reported in Fig.4. It can be shown that
equation $\theta _{e}(\lambda )=0$ is satisfied at $z=z_{cr}=1/15$.

\begin{figure}[th]
\vspace*{-5mm}
\centerline{\centerline{\psfig{figure=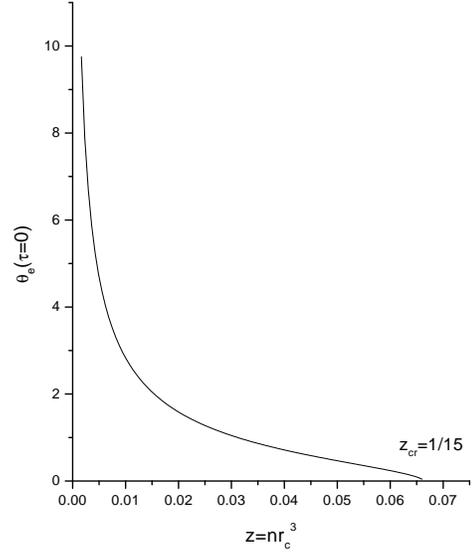,width=0.8\columnwidth}}}
\caption{Concentrational dependence of $\theta _e$ at zeroth temperature.}
\end{figure}

Let us finally show that para-glass phase cannot be realized
in the ordered dielectric. To do so we will demonstrate
the absence of "Griffiths tails" for
nondilute system. This corresponds to mean field limit ($z \to \infty$)
in random field method. In this limiting case we have from (\ref{qe5})
\begin{equation}
L_{MF}=\tanh \left( \frac{h+L_{MF}}{\tau }\right) .  \label{qe6}
\end{equation}
This equation is indeed the mean field equation for order parameter in
Ising model. In this case $\tau _{cMF}=1$.
Substitution (\ref{qe9}) to (\ref{qe6}) gives

\begin{eqnarray}
L_{1MF} &=&\frac{\sinh (2L_{1MF}/\tau )}{2\left( \sinh ^{2}%
\left( L_{1MF}/\tau \right) +\cos
^{2}\left( \theta +L_{2MF}/\tau \right) \right) },  \nonumber \\
L_{2MF} &=&\frac{\sin (2\left( \theta +L_{2MF}/\tau \right) )}{2\left( \sinh
^{2}\left( L_{1MF}/\tau \right) +\cos ^{2}%
\left( \theta +L_{2MF}/\tau \right) \right) }.
\label{qe13}
\end{eqnarray}
To calculate $\theta _{e} (\tau )$ in this case we should put
$L_{1MF}=0$ in (\ref{qe10}). This after some algebra yields

\begin{equation}
\theta _{e}=\arccos \sqrt{Q}-\sqrt{Q(1-Q)},Q=\frac{1}{\tau }.  \label{qe17}
\end{equation}
The dependences $\rho (\theta)$ and $\theta _{e} (\tau )$
for nondilute system are shown in Fig.2 and 3 (curves, labeled "MF")
respectively. It is seen the absence of "tails" in $\rho (\theta)$. This
shows, that random fields (via their distribution function $f(E,L)$) play
major role in Griffiths phase formation.

In the present paper we have shown the existence of Griffiths phase,
inherent to disordered Ising magnets, in the dielectrics with concentrational
crossover between dipole glass and ferroelectricity. We call this phase
"para-glass" because it combines the paraelectric phase features (existence
of YL edge $\theta _e$) and "disorder features" (tails in $\rho (\theta)$).
This phase resembles very much superparamagnetic phase, peculiar for many
spin glasses.

Since physical properties in glass and ferro-glass phases
of aforementioned substances are different from those in para-glass
(Griffiths) phase, we can expect some peculiarities of latter phase,
which might be helpful in its experimental observation. One of such
features may be nonexponential relaxation of polarization, different from
that in glass and ferro-glass phases. Its investigation is interesting
problem and can be done within above formalism.

Unfortunately, we don't
know the direct experimental observation of the Griffiths phase
in the aforementioned substances. For the description
of possible (direct or indirect) experiments
our approach can be easily extended to the case of nongaussian
distribution function of random fields as well as for tunneling of the
aforementioned off-center impurities between their permissible orientations
(i.e. between minima of their multi-well potentials) in the host dielectric.

\end{document}